\documentclass[italian,english]{article}
\usepackage[T1]{fontenc}
\usepackage[latin1]{inputenc}
\usepackage{color}
\usepackage{graphicx}
\usepackage{amssymb}

\makeatletter

 \newcommand{\lyxaddress}[1]{
   \par {\raggedright #1 
   \vspace{1.4em}
   \noindent\par}
 }

\usepackage{babel}
\makeatother
\begin{document}

\title{\textbf{A solution of linearized Einstein field equations in vacuum
used for the detection of the stochastic background of gravitational
waves}}

\author{\textbf{Christian Corda}}

\maketitle

\lyxaddress{\begin{center}INFN - Sezione di Pisa and Università di Pisa, Via
F. Buonarroti 2, I - 56127 PISA, Italy\end{center}}

\lyxaddress{\begin{center}\textit{E-mail address:} \textcolor{blue}{christian.corda@ego-gw.it} \end{center}}

\begin{abstract}
A solution of linearized Einstein field equations in vacuum is given
and discussed. First it is shown that, computing from our particular
metric the linearized connections, the linearized Riemann tensor and
the linearized Ricci tensor, the linearized Ricci tensor results equal
to zero. Then the effect on test masses of our solution, which is
a gravitational wave, is discussed. In our solution test masses have
an apparent motion in the direction of propagation of the wave, while
in the transverse direction they appear at rest. In this way it is
possible to think that gravitational waves would be longitudinal waves,
but, from careful investigation of this solution, it is shown that
the tidal forces associated with gravitational waves act along the
directions orthogonal to the direction of propagation of waves. The
computation is first made in the long wavelengths approximation (wavelength
much larger than the linear distances between test masses), then the
analysis is generalized to all gravitational waves.

In the last sections of this paper it is shown that the frequency
dependent angular pattern of interferometers can be obtained from
our solution and the total signal seen from an interferometer for
the stochastic background of gravitational waves is computed.
\end{abstract}

\lyxaddress{PACS numbers: 04.80.Nn, 04.30.Nk, 04.30.-w}

\section{Introduction}

The design and construction of a number of sensitive detectors for
gravitational waves (GWs) is underway today. Some laser interferometers
like the Virgo detector, being built in Cascina, near Pisa by a joint
Italian-French collaboration, the GEO 600 detector, being built in
Hanover, Germany by a joint Anglo-Germany collaboration, the two LIGO
detectors, being built in the United States (one in Hanford, Washington
and the other in Livingston, Louisiana) by a joint Caltech-Mit collaboration,
and the TAMA 300 detector, being built near Tokyo, Japan, are going
to become operative in the next years. Many bar detectors are currently
in operation too, and several interferometers and bars are in a phase
of planning and proposal stages (for the current status of gravitational
waves experiments see \cite{key-1,key-2}).

The results of these detectors will have a fundamental impact on astrophysics
and gravitation physics. There will be lots of experimental data to
be analyzed, and theorists will be forced to interact with lots of
experiments and data analysts to extract the physics from the data
stream.

Detectors for GWs will also be important to verify that GWs only change
distances perpendicular to their direction of propagation and to confirm
or ruling out the physical consistency of General Relativity or of
any other theory of gravitation \cite{key-3,key-4,key-5}.

The response of interferometers to GWs has been analyzed in lots of
works in literature especially in the so called transverse-traceless
(TT) solution of linearized Einstein field equations (LEFEs) in vacuum.
In this paper a different solution of LEFEs in vacuum is given and
discussed. First it is shown that, by computing from our particular
metric the linearized connections, the linearized Riemann tensor and
the linearized Ricci tensor (for details about linearized quantities
see \cite{key-6}), the linearized Ricci tensor results equal to zero
(i.e. our metric is solution of LEFEs). Then the effect on test masses
of this solution, which is a GW, is discussed. In our solution test
masses have an apparent motion in the direction of propagation of
the wave, while in the transverse direction they appear at rest. In
this way it is possible to think that GWs would be longitudinal waves,
but, from careful investigation of our solution, it is shown that
the tidal forces associated with GWs act along the directions orthogonal
to the direction of propagation of waves. The computation is first
made in the long wavelengths approximation (wavelength much larger
than the linear distances between test masses i.e. under this assumption
the amplitude of the GW, $h$, can be considered ''frozen'' at a
value $h_{0}$), then the analysis is extended to all GWs using a
generalization to our solution of the analysis which has been used
for scalar waves in \cite{key-5} and for tensorial waves in \cite{key-7}.

At the end of this paper it is shown that, from our solution, the
angular pattern of interferometers can be obtained with a further
generalization of the analysis of \cite{key-5,key-7} and the total
signal seen from an interferometer for the stochastic background of
GWs is computed.

\section{A solution of linearized Einstein field equations in vacuum.}

Starting from the metric (we work with $c=1$ and $\hbar=1$ in this
paper)

\begin{equation}
ds^{2}=[1+h(t-z)](-dt^{2}+dx^{2}+dz^{2})+[1-h(t-z)]dy^{2},\label{eq: metrica + 3}\end{equation}
where $h(t-z)\ll1$ is a perturbation of the flat Lorentz-Minkowski
background, let us call $\widetilde{R}_{\mu\nu\rho\sigma}$ and $\widetilde{R}_{\mu\nu}$
the linearized quantity which correspond to $R_{\mu\nu\rho\sigma}$
and $R_{\mu\nu}$. Computing them to first order in a general perturbation
$h_{\mu\nu}$, it is simple to see that the linearized Riemann tensor
is given by \cite{key-6}:

\begin{equation}
\widetilde{R}_{\mu\nu}=\frac{1}{2}(h_{\mu,\nu\alpha}^{\alpha}+h_{\nu,\mu\alpha}^{\alpha}-h_{\mu\nu,\alpha}^{\alpha}-h_{,\mu\nu}).\label{eq:Einstein2}\end{equation}

Then, because using the metric (\ref{eq: metrica + 3}) it is

\begin{equation}
h_{00}=h_{11}=h_{33}=h,\textrm{ }h_{22}=-h\label{eq: perturbazione}\end{equation}

and the other components are equal to zero, it is simple to show that
the linearized field equations

\begin{equation}
\widetilde{R}_{\mu\nu}=0,\label{eq:Einstein}\end{equation}

are satisfed.

Thus it is possible to say that the metric (\ref{eq: metrica + 3})
is a solution of LEFEs in vacuum which describes a gravitational wave
propagating in the $z+$ direction. In particular the metric (\ref{eq: metrica + 3})
is a solution for the $+$ polarization of the gravitational wave,
in fact equation (\ref{eq: metrica + 3}) can be obteined directly
from the $+$ polarization of the TT solution \cite{key-6,key-7,key-8,key-9}
with the substitution

\begin{equation}
\begin{array}{ccc}
x & \rightarrow & x\\
\\y & \rightarrow & y\\
\\z & \rightarrow & z+\frac{1}{2}H(t-z)\\
\\t & \rightarrow & t-\frac{1}{2}H(t-z),\end{array}\label{eq: transf}\end{equation}

where

\begin{equation}
H(t-z)\equiv\int_{-\infty}^{t-z}h(v)dv.\label{eq: def H zero}\end{equation}

Now we discuss the effect on test masses of this solution. Equation
(\ref{eq: metrica + 3}) can be rewritten as\begin{equation}
(\frac{dt}{d\tau})^{2}-(\frac{dx}{d\tau})^{2}-(\frac{dz}{d\tau})^{2}=\frac{1}{1+h}+\frac{1-h}{1+h}(\frac{dy}{d\tau})^{2}\label{eq: Ch2}\end{equation}

where $\tau$ is the proper time of the test masses.

To derive the geodesic equation of motion for test masses (i.e. the
beam-splitter and the mirrors of an interferometer) the equation

\begin{equation}
\frac{du_{i}}{d\tau}-\frac{1}{2}\frac{\partial g_{kl}}{\partial x^{i}}u^{k}u^{l}=0,\label{eq: Landau geodesic}\end{equation}

can be used \cite{key-8}.

Thus, from the metric (\ref{eq: metrica + 3}) we obtain

\begin{equation}
\begin{array}{ccc}
\frac{d^{2}x}{d\tau^{2}} & = & 0\\
\\\frac{d^{2}y}{d\tau^{2}} & = & 0\\
\\\frac{d^{2}t}{d\tau^{2}} & = & \frac{1}{2(1+h)}\partial_{t}(1+h)[(\frac{dt}{d\tau})^{2}-(\frac{dx}{d\tau})^{2}-(\frac{dz}{d\tau})^{2}]-\frac{1}{2}\partial_{t}(1-h)(\frac{dy}{d\tau})^{2}\\
\\\frac{d^{2}z}{d\tau^{2}} & = & -\frac{1}{2(1+h)}\partial_{z}(1+h)[(\frac{dt}{d\tau})^{2}-(\frac{dx}{d\tau})^{2}-(\frac{dz}{d\tau})^{2}]+\frac{1}{2}\partial_{z}(1-h)(\frac{dy}{d\tau})^{2}.\end{array}\label{eq: geodetiche Corda*}\end{equation}

The first and the second of eqs. (\ref{eq: geodetiche Corda*}) can
be immediately integrated obtaining

\begin{equation}
\frac{dx}{d\tau}=C_{1}=const.\label{eq: integrazione x}\end{equation}

\begin{equation}
\frac{dy}{d\tau}=C_{2}=const.\label{eq: integrazione x}\end{equation}

Assuming that test masses are at rest initially we get $C_{1}=C_{2}=0$.
Thus, even if the GW arrives at test masses, there is no motion of
test masses within the $x-y$ plane. This could be directly understood
from eq. (\ref{eq: metrica + 3}) because the absence of the $x$
and of the $y$ dependences in the metric implies that test masses
momentum in these directions (i.e. $C_{1}$ and $C_{2}$ respectively)
is conserved. This results, for example, from the fact that in this
case the $x$ and $y$ coordinates do not esplicity enter in the Hamilton-Jacobi
equation for a test mass in a gravitational field (see \cite{key-8}
for details).

Now eq. (\ref{eq: Ch2}) reads

\begin{equation}
(\frac{dt}{d\tau})^{2}-(\frac{dz}{d\tau})^{2}=\frac{1}{1+h}.\label{eq: Ch3}\end{equation}

In this way, eqs. (\ref{eq: geodetiche Corda*}) begin

\begin{equation}
\begin{array}{ccc}
\frac{d^{2}x}{d\tau^{2}} & = & 0\\
\\\frac{d^{2}y}{d\tau^{2}} & = & 0\\
\\\frac{d^{2}t}{d\tau^{2}} & = & \frac{1}{2}\frac{\partial_{t}(1+h)}{(1+h)^{2}}\\
\\\frac{d^{2}z}{d\tau^{2}} & = & -\frac{1}{2}\frac{\partial_{z}(1+h)}{(1+h)^{2}}.\end{array}\label{eq: geodetiche Corda}\end{equation}

Now it will be shown that, in presence of a GW, there will be motion
of test masses in the $z$ direction which is the direction of the
propagating wave. An analysis of eqs. (\ref{eq: geodetiche Corda})
shows that, to simplify equations, the retarded and advanced time
coordinates ($v,w$) can be introduced:

\begin{equation}
\begin{array}{c}
v=t-z\\
\\w=t+z.\end{array}\label{eq: ret-adv}\end{equation}

From the third and the fourth of eqs. (\ref{eq: geodetiche Corda})
we have

\begin{equation}
\frac{d}{d\tau}\frac{dv}{d\tau}=\frac{\partial_{w}[1+h(v)]}{[1+h(v)]^{2}}=0.\label{eq: t-z t+z}\end{equation}

Thus we obtain

\begin{equation}
\frac{dv}{d\tau}=\alpha,\label{eq: t-z}\end{equation}

where $\alpha$ is an integration constant. From eqs. (\ref{eq: Ch3})
and (\ref{eq: t-z}), we get

\begin{equation}
\frac{dw}{d\tau}=\frac{\beta}{1+h}\label{eq: t+z}\end{equation}

where $\beta\equiv\frac{1}{\alpha}$, and

\begin{equation}
\tau=\beta v+\gamma,\label{eq: tau}\end{equation}

where the integration constant $\gamma$ correspondes simply to the
retarded time coordinate translation $v=t-z$. Thus it can be put
equal to zero without loss of generality. Now let us see what is the
meaning of the other integration constant $\beta$. From eqs. (\ref{eq: t-z})
and (\ref{eq: t+z}) the equation for $z$ can be written:

\begin{equation}
\frac{dz}{d\tau}=\frac{1}{2\beta}(\frac{\beta^{2}}{1+h}-1).\label{eq: z}\end{equation}

When it is $h=0$ (i.e. before the GW arrives at the test masses)
eq. (\ref{eq: z}) becomes\begin{equation}
\frac{dz}{d\tau}=\frac{1}{2\beta}(\beta^{2}-1).\label{eq: z ad h nullo}\end{equation}

But this is exactly the initial velocity of the test mass, thus we
have to choose $\beta=1$ because we suppose that test masses are
at rest initially. This also imply $\alpha=1$.

To find the motion of a test mass in the $z$ direction from eq. (\ref{eq: tau})
we have $d\tau=dv$, while from eq. (\ref{eq: t+z}) we have $dw=\frac{d\tau}{1+h}$.
Because it is $z=\frac{w-v}{2}$ we obtain

\begin{equation}
dz=\frac{1}{2}(\frac{d\tau}{1+h}-dv),\label{eq: dz}\end{equation}

which can be integrated as

\begin{equation}
\begin{array}{c}
z=z_{0}+\frac{1}{2}\int(\frac{dv}{1+h}-dv)=\\
\\=z_{0}-\frac{1}{2}\int_{-\infty}^{t-z}\frac{h(v)}{1+h(v)}dv,\end{array}\label{eq: moto lungo z}\end{equation}

where $z_{0}$ is the initial position of the test mass. Now the displacement
of the test mass in the $z$ direction can be written as

\begin{equation}
\begin{array}{c}
\Delta z=z-z_{0}=-\frac{1}{2}\int_{-\infty}^{t-z_{0}-\Delta z}\frac{h(v)}{1+h(v)}dv\\
\\\simeq-\frac{1}{2}\int_{-\infty}^{t-z_{0}}\frac{h(v)}{1+h(v)}dv.\end{array}\label{eq: spostamento lungo z}\end{equation}

Our results can be also rewritten in function of the time coordinate
$t$:

\begin{equation}
\begin{array}{ccc}
x(t) & = & x_{0}\\
\\y(t) & = & y_{0}\\
\\z(t) & = & z_{0}-\frac{1}{2}\int_{-\infty}^{t-z_{0}}\frac{h(v)}{1+h(v)}d(v)\\
\\\tau(t) & = & t-z(t).\end{array}\label{eq: moto sol Corda}\end{equation}

Now let us reasume what happens in our solution: it has been shown
that in the $x-y$ plane an inertial test mass initially at rest remains
at rest throughout the entire passage of the GW, while in the $z$
direction an inertial test mass initially at rest has a motion during
the passage of the GW. Thus it could appear that the solution (\ref{eq: metrica + 3})
for GWs has a longitudinal effect and does not have a transversal
one, but the situation is different as it will be shown in the following
analysis.

\section{Analysis in the long wavelenghts approximation}

We have to clarify the use of words {}`` at rest'' : we want to
mean that the coordinates of test masses do not change in the presence
of the GW in the $x-y$ plane \cite{key-5,key-6,key-7,key-9}, but
it will be shown that the proper distance between the beam-splitter
and the mirror of our interferometer changes even though their coordinates
remain the same. On the other hand, it will be also shown that the
proper distance between the beam-splitter and the mirror of our interferometer
does not change in the $z$ direction even if their coordinates change
in the solution (\ref{eq: metrica + 3}). 

A good way to analyze variations in the proper distance (time) is
by means of {}``bouncing photons'' : a photon can be launched from
the beam-splitter to be bounced back by the mirror (see \cite{key-5,key-7}
and figure 1). 

\begin{figure}
\includegraphics{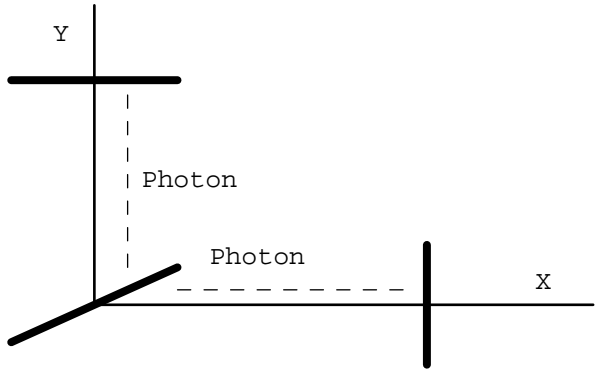}

\caption{photons can be launched from the beam-splitter to be bounced back
by the mirror}
\end{figure}

In this section we only deal with the case in which the frequency
$f$ of the GW is much smaller than $\frac{1}{T_{0}}=\frac{1}{L_{0}}$,
where $2T_{0}=2L_{0}$ is the total round-trip time of the photon
in absence of the GW. The analysis will be generalized to all frequencies
in the next section.

We assume that test masses are located along the $x$ axis and the
$z$ axis of the coordinate system. In this case the $y$ direction
can be neglected because the absence of the $y$ dependence in the
metric (\ref{eq: metrica + 3}) implies that photon momentum in this
direction is conserved \cite{key-5,key-7,key-8} and the interval
can be rewritten in the form

\begin{equation}
ds^{2}=[1+h(t-z)](-dt^{2}+dx^{2}+dz^{2}).\label{eq: metrica + 4}\end{equation}

Let us start by considering the interval for a photon which propagates
in the $x$ axis. We note that photon momentum in the $z$ direction
is not conserved, for the $z$ dependence in eq. (\ref{eq: metrica + 3})
\cite{key-5,key-7,key-8}. Thus photons launched in the $x$ axis
will deflect out of this axis. But here this effect can be neglected
because the photon deflection into the $z$ direction will be at most
of order $h$ (see \cite{key-7}). Then, to first order in $h$, the
$dz^{2}$ term can be neglected. Thus, eq. (\ref{eq: metrica + 4})
can be rewritten as

\begin{equation}
ds^{2}=(1+h)(-dt^{2})+(1+h)dx^{2}.\label{eq: metrica + 3 lungo x}\end{equation}

The condition for null geodesics ($ds^{2}=0$) for photons gives 

\begin{equation}
\frac{dx_{photon}}{dt}=\pm1\Rightarrow x_{photon}=const\pm t.\label{eq: null geodesic}\end{equation}

In our solution the $x$ coordinates of the beam-splitter and the
mirrors are unaffected by the passage of the GW (see the first of
eqs. (\ref{eq: moto sol Corda}) ), then, from eq. (\ref{eq: null geodesic}),
it is possible to see that the interval, in coordinate time $t$,
that the photon takes for run one round trip in the $x$ arm of the
interferometer is

\begin{equation}
T=2L_{0}\label{eq:  tempo doppio}\end{equation}

(i.e. the photon leaves the beam-splitter at $t=0$ and returns a
$t=T$). But this quantity is not invariant under coordinate transformations,
and we have to work in terms of the beam-splitter proper time which
misures the physical lenght of the arms. In this way, calling $\tau(t)$
and $z_{b}(t)$ the proper time and $z$ coordinate of the beam-splitter
at time coordinate $t$ with initial condition $z_{b}(-\infty)=0$,
from eqs. (\ref{eq: moto sol Corda}) it is possible to obtain

\begin{equation}
\begin{array}{ccc}
z_{b}(t) & = & -\frac{1}{2}\int_{-\infty}^{t-z_{b}(t)}\frac{h(v)}{1+h(v)}dv\\
\\\tau(t) & =t & +\frac{1}{2}\int_{-\infty}^{t-z_{b}(t)}\frac{h(v)}{1+h(v)}dv.\end{array}\label{eq: bm lungo z}\end{equation}

Thus, calling $\tau_{x}$ the proper time interval that the photon
takes to run a round-trip in the $x$ arm, we have

\begin{equation}
\begin{array}{c}
\tau_{x}=\tau(T)-\tau(0)=T+\frac{1}{2}\int_{-z_{b}(0)}^{t-z_{b}(t)}\frac{h(v)}{1+h(v)}dv\simeq\\
\\\simeq T+\frac{1}{2}h_{0}[T+z_{b}(0)-z_{b}(T)]\simeq\\
\\\simeq2L_{0}(1+\frac{1}{2}h_{0}).\end{array}\label{eq: tempo proprio lungo x}\end{equation}

In the above computation eq. (\ref{eq:  tempo doppio}) has been used
and, considering only the first order in $h$ with $h\ll1$, the field
$h$ has been also considered {}``frozen'' at a fixed value $h_{0}$.
We note that $z_{b}(0)-z_{b}(T)$ is second order in $h_{0}$.

Now let us consider the $z$ direction: the $x$ direction can be
neglected because the absence of the $x$ dependence in the metric
(\ref{eq: metrica + 3}) implies that photon momentum in this direction
is conserved \cite{key-5,key-7,key-8}. From eq. (\ref{eq: metrica + 4})
it is now:

\begin{equation}
ds^{2}=(1+h)(-dt^{2})+(1+h)dz^{2},\label{eq: metrica puramente piu di Corda lungo z}\end{equation}

and the condition for null geodesics ($ds^{2}=0$) for photons gives 

\begin{equation}
\frac{dz_{photon}}{dt}=\pm1\Rightarrow z_{photon}=const\pm t.\label{eq: null geodesic2}\end{equation}
Supposing that the photon leaves the beam splitter at $t=0$ let us
ask: how much time does the photon need to arrive at the mirror in
the $z$ axis? Calling $T_{1}$ this time we need the condition

\begin{equation}
z_{b}(0)+T_{1}=z_{m}(T_{1}),\label{eq: z.m}\end{equation}

where $z_{m}(t)$ is the $z$ coordinate of the mirror in the $z$
axis at coordinate time $t$ with $z_{m}(-\infty)=L_{0}$. In the
same way, when returning from the mirror the photon arrives again
at the beam-splitter at $t=T_{z}=T_{1}+T_{2}$, then

\begin{equation}
z_{m}(T_{1})-T_{2}=z_{b}(T_{z}).\label{eq: z.m2}\end{equation}

Subtracting eq. (\ref{eq: z.m2}) from eq. (\ref{eq: z.m}) we obtain 

\begin{equation}
T_{z}=T_{1}+T_{2}=[z_{m}(T_{1})-z_{b}(0)]+[z_{m}(T_{1})-z_{b}(T_{z})].\label{eq: tempo}\end{equation}

It is known from eq. (\ref{eq: moto sol Corda}) that the equations
of motion for $z_{b}$ and $z_{m}$ are:

\begin{equation}
\begin{array}{ccc}
z_{m}(t) & = & L_{0}-\frac{1}{2}\int_{-\infty}^{t-z_{m}(t)}\frac{h(v)}{1+h(v)}dv\\
\\z_{b}(t) & = & -\frac{1}{2}\int_{-\infty}^{t-z_{b}(t)}\frac{h(v)}{1+h(v)}dv,\end{array}\label{eq: bm lungo z}\end{equation}

and, substituing them in eq. (\ref{eq: tempo}), we get

\begin{equation}
T_{z}=2L_{0}-\frac{1}{2}\int_{-z_{b}(0)}^{T_{1}-z_{m}(T_{1})}\frac{h(v)}{1+h(v)}dv-\frac{1}{2}\int_{T_{z}-z_{b}(T_{z})}^{T_{1}-z_{m}(T_{1})}\frac{h(v)}{1+h(v)}dv.\label{eq: tempo lungo z}\end{equation}

From eq. (\ref{eq: z.m}) we see that the first integral in eq. (\ref{eq: tempo lungo z})
is zero. The second integral is simple to compute if the GW is considered
frozen at a value $h_{0}$. To first order in this value it is

\begin{equation}
\begin{array}{c}
-\frac{1}{2}\int_{T_{z}-z_{b}(T_{z})}^{T_{1}-z_{m}(T_{1})}\frac{h(v)}{1+h(v)}dv\simeq-\frac{1}{2}h_{0}[T_{1}-z_{m}(T_{1})-T_{z}+z_{b}(T_{z})]\simeq\\
\\\simeq-\frac{1}{2}h_{0}(L_{0}-L_{0}-2L_{0})=+\frac{1}{2}h_{0}2L_{0}.\end{array}\label{eq: int2}\end{equation}

In this way eq. (\ref{eq: tempo lungo z}) becomes

\begin{equation}
T_{z}=(1+\frac{1}{2}h_{0})2L_{0}.\label{eq: tempo lungo z2}\end{equation}

Then, calling $\tau_{z}$ the proper time interval that the photon
takes to run a round-trip in the $z$ arm, with the same way of thinking
which leaded to eq. (\ref{eq: tempo proprio lungo x}), we obtain

\begin{equation}
\begin{array}{c}
\tau_{z}=\tau(T_{z})-\tau(0)=T_{z}+\frac{1}{2}\int_{-z_{b}(0)}^{t-z_{b}(t)}\frac{h(v)}{1+h(v)}dv\simeq\\
\\\simeq T_{z}+\frac{1}{2}h_{0}[T_{z}+z_{b}(0)-z_{b}(T)]\simeq\\
\\\simeq T_{z}(1-\frac{1}{2}h_{0})\simeq2L_{0}.\end{array}\label{eq: tempo proprio lungo z}\end{equation}

Thus, from eqs. (\ref{eq: tempo proprio lungo x}) and (\ref{eq: tempo proprio lungo z})
it is shown that there is a variation of the proper distance in the
$x$ direction (transversal effect of the GW), while there is not
a variation of the proper distance in the $z$ direction (no longitudinal
effect).

\section{Generalization of the analysis}

Now the previous result will be generalized to all the frequencies,
with an analysis that, with a trasform of the time coordinate to the
proper time, generalizes to our solution the analysis of \cite{key-7}
, where the analysis was made using the TT solution of LEFEs in vacuum.
In this way it will also be obtained the response function of the
interferometer for our solution.

Let us start with the $x$ arm of the interferometer. The condition
of null geodesic (\ref{eq: null geodesic}) can also be rewrite in
this way:

\begin{equation}
dt^{2}=dx^{2}.\label{eq: metrica puramente piu' di Corda lungo x 2}\end{equation}

Thus, in this case, the analysis of \cite{key-7} cannot be used starting
directly from the condition of null geodesic. In fact the metric (\ref{eq: metrica + 3 lungo x})
is different from the metric of eq. (3) in \cite{key-7}. In \cite{key-7}
the author used the condition of null geodesic to obtain the coordinate
velocity of the photon which was used for calculations of the photon
propagation times between the test masses (eq. (4) in \cite{key-7}).
But in equation (\ref{eq: metrica puramente piu' di Corda lungo x 2})
appears that the coordinate velocity of the photon is equal to the
speed of light in our solution. Then let us ask which is the important
difference between our metric (\ref{eq: metrica + 3}) and the TT
metric analyzed in \cite{key-7}. The answer is that the TT metric
defines a {}``synchrony coordinate system'', a coordinate system
in which the time coordinate $t$ is exactly the proper time (about
the synchrony coordinate system see Cap. (9) of \cite{key-8} and
Section 4 of \cite{key-5}). In the solution (\ref{eq: metrica + 3})
$t$ is only a time coordinate. We know that the rate $d\tau$ of
the proper time is related to the rate $dt$ of the time coordinate
from \cite{key-5,key-8}

\begin{equation}
d\tau^{2}=g_{00}dt^{2}.\label{eq: relazione temporale}\end{equation}

Only with the aid of the time transform (\ref{eq: relazione temporale})
the analysis of \cite{key-7} can be applied to our solution.

From eq. (\ref{eq: metrica + 3 lungo x}) it is $g_{00}=(1+h)$. Then,
using eq. (\ref{eq: metrica puramente piu' di Corda lungo x 2}),
we obtain

\begin{equation}
d\tau^{2}=(1+h)dx^{2},\label{eq: relazione spazial-temporale}\end{equation}

which gives 

\begin{equation}
d\tau=\pm(1+h)^{\frac{1}{2}}dx.\label{eq: relazione temporale 2}\end{equation}

Now it will be shown that the analysis of \cite{key-7} works in our
case too. This is an analysis parallel to the one used in Section
4 of \cite{key-5}.

In eqs. (\ref{eq: moto sol Corda}) it is shown that the coordinates
of the beam-splitter $x_{b}=l$ and of the mirror $x_{m}=l+L_{0}$
do not change under the influence of the GW in our solution, thus
the proper duration of the forward trip can be found as

\begin{equation}
\tau_{1}(t)=\int_{l}^{L_{0}+l}[1+h(t)]^{\frac{1}{2}}dx.\label{eq: relazione temporale 2}\end{equation}

To first order in $h$ this integral can be approximated with

\begin{equation}
\tau_{1}(t)=T_{0}+\frac{1}{2}\int_{l}^{L_{0}+l}h(t')dx\label{eq: durata volo andata approssimata in Corda x}\end{equation}

where

\begin{center}$t'=t-(l+L_{0}-x)$.\end{center}

In the last equation $t'$ is the retardation time (i.e. $t$ is the
time at which the photon arrives in the position $l+L_{0}$, so $l+L_{0}-x=t-t'$
\cite{key-5,key-7}).

In the same way we have for the proper duration of the return trip, 

\begin{equation}
\tau_{2}(t)=T_{0}+\frac{1}{2}\int_{l+L_{0}}^{l}h(t')(-dx),\label{eq: durata volo ritorno approssimata in Corda x}\end{equation}

where now 

\begin{center}$t'=t-(x-l)$\end{center}

is the retardation time and

\begin{center}$T_{0}=L_{0}$ \end{center}

is the transit proper time of the photon in the absence of the GW,
which also corresponds to the transit coordinate time of the photon
in the presence of the GW (see eq. (\ref{eq: metrica puramente piu' di Corda lungo x 2})).

Thus the round-trip proper time will be the sum of $\tau_{2}(t)$
and $\tau_{1}(t-T_{0})$. Then, to first order in $h$, the proper
duration of the round-trip will be

\begin{equation}
\tau_{r.t.}(t)=\tau_{1}(t-T_{0})+\tau_{2}(t).\label{eq: durata round trip}\end{equation}

From eqs. (\ref{eq: durata volo andata approssimata in Corda x})
and (\ref{eq: durata volo ritorno approssimata in Corda x}) it is
immediately shown that deviations of this round-trip proper time (i.e.
proper distance) from its imperurbated value are given by

\begin{equation}
\delta\tau(t)=\frac{1}{2}\int_{l}^{L_{0}+l}[h(t-2T_{0}+x-l)+h(t-x+l)]dx.\label{eq: variazione temporale in sol Corda}\end{equation}

Eq. (\ref{eq: variazione temporale in sol Corda}) generalizes eq.
(\ref{eq: tempo proprio lungo x}) which was derived in the low frequencies
approximation. The signal seen from the arm in the $x$ axis can be
also defined like

\begin{equation}
\frac{\delta\tau(t)}{T_{0}}\equiv\frac{1}{2T_{0}}\int_{l}^{L_{0}+l}[h(t-2T_{0}+x-l)+h(t-x+l)]dx.\label{eq: signal}\end{equation}

Now the analysis will be transled in the frequency domain using the
Fourier transform of the field $h$ defined by

\begin{equation}
\tilde{h}(\omega)=\int_{-\infty}^{\infty}dt\textrm{ }h(t)\exp(i\omega t).\label{eq: trasformata di fourier1}\end{equation}

With the definition (\ref{eq: trasformata di fourier1}), from eq.
(\ref{eq: signal}) we have

\begin{equation}
\frac{\delta\tilde{\tau}(\omega)}{T_{0}}=\Upsilon(\omega)\tilde{h}(\omega),\label{eq: fourier in sol Corda}\end{equation}

where $\Upsilon(\omega)$ is the response of the $x$ arm of our interferometer
to GWs:

\begin{equation}
\Upsilon(\omega)=\frac{\exp(2i\omega T_{0})-1}{2i\omega T_{0}},\label{eq: risposta in sol Corda}\end{equation}

which is computated in lots of works in literature. 

Now let us see what happens in the $z$ coordinate (see figure 2).
\begin{figure}
\includegraphics{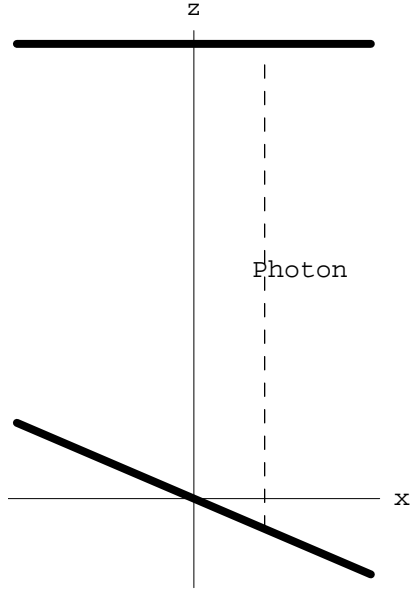}

\caption{the beam splitter and the mirror are located in the direction of
the incoming GW}
\end{figure}

Eq. (\ref{eq: metrica puramente piu di Corda lungo z}) and the condition
$ds^{2}=0$ for null geodesics also give

\begin{equation}
dz=\pm dt.\label{eq: metrica puramente piu di Corda lungo z 2}\end{equation}

But, from the last of eqs. (\ref{eq: moto sol Corda}) the proper
time can be written like

\begin{equation}
d\tau(t)=dt-dz,\label{eq: tempo proprio lungo z in Corda}\end{equation}

and, combining eq. (\ref{eq: metrica puramente piu di Corda lungo z 2})
with eq. (\ref{eq: tempo proprio lungo z in Corda}), we get

\begin{equation}
d\tau(t)=dt\mp dt.\label{eq: tempo proprio lungo z in Corda 2}\end{equation}

Thus we obtain now

\begin{equation}
\tau_{1}(t)=0\label{eq:  tempo di propagazione andata sol Corda lungo z}\end{equation}

for the forward trip

and

\begin{equation}
\tau_{2}(t)=\int_{0}^{T_{0}}2dt=2T_{0}\label{eq:  tempo di propagazione ritorno sol Corda  lungo z}\end{equation}

for the return trip. Then

\begin{equation}
\tau(t)=\tau_{1}(t)+\tau_{2}(t)=2T_{0}.\label{eq: tempo proprio totale lungo z in Corda}\end{equation}

Thus there is not longitudinal effect (i.e. $\delta\tau=\delta L_{0}=0$).
This is a direct conseguence of the fact that a GW propagates at the
speed of light. In this way in the forward trip the photon travels
at the same speed of the GW and its proper time is equal to zero (eq.
(\ref{eq:  tempo di propagazione andata sol Corda lungo z})), while
in the return trip the photon travels \textbf{against} the GW and
its proper time redoubles (eq. (\ref{eq:  tempo di propagazione ritorno sol Corda  lungo z})).

\section{The low-frequencies angular pattern of interferometers in the TT
solution of linearized Einstein field equations in vacuum}

Very important in a cosmological scenario is the concept of stochastic
background of gravitational waves \cite{key-9,key-10,key-11,key-12,key-13}.

Roughly speaking these are GWs that are produced by a very large number
of weak, independent and unresolved sources \cite{key-9,key-10,key-12,key-13}.
A stochastic background of GWs could be the result of processes that
took place in a time very close to the Planck era \cite{key-9,key-11,key-12,key-13}.
But it could be due to more recent processes too. An example can be
radiation from many unresolved binary systems like neutron stars,
white dwarfs and black holes. These more recent contributions could
overwhelm the primordial parts of the background. In this context
we know that, in any case, the properties of the radiation will be
strongly dependent upon the source. For example, a stochastic background
of relic GWs is expected to be isotropic, while gravitational radiation
derived by binary sistems in our galaxy would be highly anisotropic.
Thus we have to see the output of our detectors before taking a decision
between these two possibilities \cite{key-10,key-12,key-13}.

In which sense sources of the stochastic background are {}``unresolved''
can be understood making an analogy with optical sources. In the study
of an optical source, somewhere in the sky, using a telescope with
a certain angular resolution, details of the source can be resolved
if the angular resolution of our telescope is smaller than the angular
size of the features of the source. In the case of Virgo (and similiar
experiments like the two LIGO), the angular size of the detector pattern
is of order $90^{\circ}$. Thus almost any source could make a significant
contribution to the detector strain for almost any orientation of
both detector and source, and the sources are unresolved. When lots
of sources that give a contribution are present, even if they are
pointlike, the resulting signal is stochastic. 

Thus, in this scenario, it is simple to understand the fundamental
importance of the angular pattern of a detector. 

Let us see what happens in a detector. The total output of the antenna
$S(t)$ is in general of the form

\begin{equation}
S(t)=s(t)+n(t),\label{eq: output detector}\end{equation}

where $n(t)$ is the noise and $s(t)$ is the contribution to the
output due to the gravitational waves.

For an interferometer with equal arms of lenght $L$ (3 kilometers
in the case of Virgo) in the $u-v$ plane, it is

\begin{equation}
s(t)=\frac{\delta L_{u}(t)-\delta L_{v}(t)}{L},\label{eq: output gravitoni}\end{equation}
where $\delta L_{u,v}$ are the displacements produced by gravitational
waves. In the TT solution of LEFEs in vacuum (this solution is historically
called transverse-traceless (TT), because the gravitational waves
have a transverse effect and are traceless \cite{key-6}) the total
perturbation of a gravitational wave propagating in the positive $\overrightarrow{z}=z\hat{\Omega}$
direction and with a wave front parallel to the $x-y$ plane \cite{key-6}
is given by

\begin{equation}
h_{\alpha\beta}(t-z)=h^{+}(t-z)e_{\alpha\beta}^{+}(\hat{\Omega})+h^{\times}(t-z)e_{\alpha\beta}^{\times}(\hat{\Omega}),\label{eq: onda generalizzata}\end{equation}

where $e_{\alpha\beta}^{+}(\hat{\Omega})$ and $e_{\alpha\beta}^{\times}(\hat{\Omega})$
are the two polarizations \cite{key-9,key-10,key-11,key-12,key-13}:

\begin{equation}
e_{\alpha\beta}^{+}=\hat{x}^{\alpha}\hat{x}^{\beta}-\hat{y}^{\alpha}\hat{y}^{\beta}\label{eq: polarizzazione piu'}\end{equation}

\begin{equation}
e_{\alpha\beta}^{\times}=\hat{x}^{\alpha}\hat{x}^{\beta}+\hat{y}^{\alpha}\hat{y}^{\beta}.\label{eq: polarizzazione per}\end{equation}

It is also known that the relation between the output $s(t)$ and
the total signal of gravitational waves $h_{ab}(t)$ in the TT solution
of LEFEs in vacuum has the form \cite{key-9,key-11,key-12,key-13}

\begin{equation}
s(t)=D^{ab}h_{ab}(t),\label{eq: legame onda-output}\end{equation}

where $D^{ab}$ is called \textit{detector tensor}. For an interferometer
with arms along the $\hat{u}$ e $\hat{v}$ directions (not necessarly
orthogonal) we have

\begin{equation}
D^{ab}\equiv\frac{1}{2}(\hat{v}^{a}\hat{v}^{b}-\hat{u}^{a}\hat{u}^{b}).\label{eq: definizione D}\end{equation}

Thus, in the case of the stochastic bacground of gravitational waves,
the equation 

\begin{equation}
\begin{array}{c}
h_{ab}(t,\overrightarrow{x})=\\
\\=\frac{1}{2\pi}\sum_{A}\int_{-\infty}^{+\infty}d\omega\int_{S^{2}}d\hat{\Omega}\tilde{h}_{A}(\omega,\hat{\Omega})\exp i\omega(t-\hat{\Omega}\cdot\overrightarrow{x})e_{ab}^{A}(\hat{\Omega})\end{array}\label{eq: fourier expansion}\end{equation}

can be used for the total signal \cite{key-9,key-10,key-11,key-12,key-13,key-14},
where 

\begin{equation}
\hat{\Omega}=\cos\phi\sin\theta\hat{x}+\sin\phi\sin\theta\hat{y}+\cos\theta\hat{z},\label{eq: Omega}\end{equation}

with

\begin{equation}
d\hat{\Omega}=d\cos\theta d\phi\label{eq:domega}\end{equation}

and\begin{equation}
e_{ab}^{A}(\hat{\Omega})\equiv[e_{ab}^{+}(\hat{\Omega}),e_{ab}^{\times}(\hat{\Omega})],\label{eq: pol}\end{equation}

are the two polarizations (\ref{eq: polarizzazione piu'}) and (\ref{eq: polarizzazione per}).

Putting $\overrightarrow{x}=0$ in our expansion (i.e. the coordinates
of the detector are in the origin of our system), we obtain

\begin{equation}
s(t)=\frac{1}{2\pi}\sum_{A}\int_{-\infty}^{+\infty}d\omega\int_{S^{2}}d\hat{\Omega}\tilde{h}_{A}(\omega,\hat{\Omega})\exp(i\omega t)D^{ab}e_{ab}^{A}(\hat{\Omega}).\label{eq: legame output-stochastic}\end{equation}

The correspondent equation in the frequency domain is also given by:

\begin{equation}
\tilde{S}(\omega)=D^{ab}\tilde{h}_{ab}(\omega),\label{eq: formula  approssimata in frequenza}\end{equation}

which can be rewritten as

\begin{equation}
\tilde{S}(\omega)=\int_{S^{2}}d\hat{\Omega}\tilde{h}_{A}(\omega,\hat{\Omega})D^{ab}e_{ab}^{A}(\hat{\Omega}).\label{eq: formula  approssimata in frequenza2}\end{equation}

The quantity

\begin{equation}
F^{A}(\Omega)\equiv D^{ab}e_{ab}^{A}(\hat{\Omega})=Tr\{ De\},\label{eq: pattern alle basse frequenze}\end{equation}

is called \textit{detector pattern}.

Thus, using eqs. (\ref{eq: polarizzazione piu'}), (\ref{eq: polarizzazione per})
and (\ref{eq: definizione D}), combined with eq. (\ref{eq: pattern alle basse frequenze})
it is simple to obtain

\begin{equation}
F^{+}(\Omega)=\frac{1}{2}(1+\cos^{2}\theta)\cos2\phi\label{eq: pattern piu'}\end{equation}

\begin{equation}
F^{\times}(\Omega)=-\cos\theta\sin2\phi\label{eq: pattern per}\end{equation}

These detector patterns for different polarizations of GWs have been
analyzed in lots of works in literature (see for example \cite{key-12,key-13,key-14,key-15}).

But there is a problem: eqs. (\ref{eq: pattern piu'}) and (\ref{eq: pattern per})
are not the general form of the detector patterns, but they are only
a good approximation for long wavelengths (i.e. the wavelength of
the wave is much larger than the linear dimension of the interferometer)
\cite{key-15}. In this approximation the detector can be considered
pointlike when a gravitational wave is arriving (i.e. we can see the
wave ''frozen'' at a value $h_{0}$).

In the next Section, with the auxilium of our solution of LEFEs in
vacuum, the exact frequency - dependent expressions for equations
(\ref{eq: pattern piu'}) and (\ref{eq: pattern per}) will be derived
in the case of an interferometer with perpendicular arms.

\section{The detector pattern in the general case in our solution.}

In the context of the potential detection of stochastic backgrounds
of GWs with interferometers in this Section it will be generalized
to all the wavelengths the concept of angular pattern for interferometers,
which is well known for the TT solution of LEFEs in vacuum and in
the assumption that the wavelength of the GWs is much larger than
the distance between the test masses (i.e. the beam-splitter and the
mirrors of the interferometer, see Section 5 and refs. \cite{key-12,key-13,key-14,key-15}). 

We emphasize that, in the analysis of the angular pattern of interferometers,
a further generalization of the analysis of \cite{key-5,key-7} will
be made. In \cite{key-7} only the simplest geometry case was considered
in the purely $+$ polarization of the wave while in \cite{key-5}
the purely scalar case was analyzed. Here it will be shown that the
analysis can be generalized for the more general geometry of the $+$
polarization and for the $\times$ polarization of GWs too.

It will be computed the variaton of the proper distance that a photon
covers to make  a round-trip from the beam-splitter to the mirror
of an interferometer \cite{key-5,key-7} with a coordinates choice
that, for the $+$ polarization gives the line element (\ref{eq: metrica + 3}),
while for the $\times$ polarization gives the line element 

\begin{equation}
ds^{2}=(-dt^{2}+dz^{2})[1+h^{\times}(t-z)]+dx^{2}+dy^{2}+2h^{\times}(t-z)dxdy,\label{eq: metrica polarizzazione per in gauge Corda}\end{equation}

which can be obteined applying the substitution (\ref{eq: transf})
to the TT solution for the $\times$ polarization.

It is simple to see that also the metric (\ref{eq: metrica polarizzazione per in gauge Corda})
satisfes LEFEs in vacuum.

Now, with a treatment which generalize to the angular dependence the
analysis of \cite{key-7} and wich is parallel to the analysis in
Section 7 of \cite{key-5}, the computation will be transled in the
frequency domain and the general frequency dependent angular patterns
of interferometers will be derived.

We start from the $+$ polarization. In this case the interval (\ref{eq: metrica + 3})
has to be considered.

But we recall that the arms of our interferometer are in the $\overrightarrow{u}$
and $\overrightarrow{v}$ directions, thus, to compute the line element
in the $\overrightarrow{u}$ and $\overrightarrow{v}$ directions,
a spatial rotation of our coordinates has to be made \cite{key-5}:

\begin{equation}
\begin{array}{ccc}
u & = & -x\cos\theta\cos\phi+y\sin\phi+z\sin\theta\cos\phi\\
\\v & = & -x\cos\theta\sin\phi-y\cos\phi+z\sin\theta\sin\phi\\
\\w & = & x\sin\theta+z\cos\theta,\end{array}\label{eq: rotazione}\end{equation}

or, in terms of the $x,y,z$ frame:

\begin{equation}
\begin{array}{ccc}
x & = & -u\cos\theta\cos\phi-v\cos\theta\sin\phi+w\sin\theta\\
\\y & = & u\sin-v\cos\phi\\
\\z & = & u\sin\theta\cos\phi+v\sin\theta\sin\phi+w\cos\theta.\end{array}\label{eq: rotazione 2}\end{equation}

In this way the GW is propagating from an arbitrary direction $\overrightarrow{r}$
to the interferometer (see figure 3). 

\begin{figure}
\includegraphics{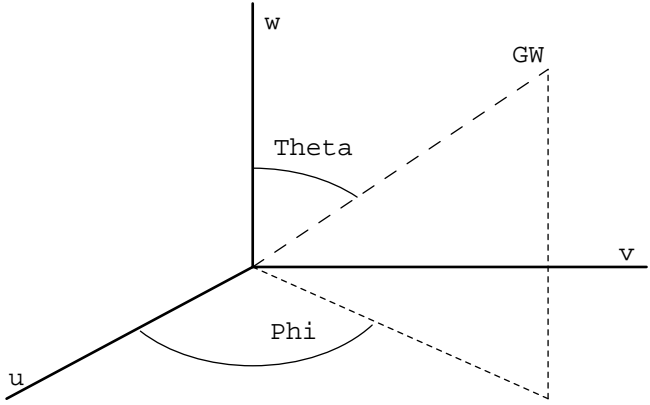}

\caption{a GW incoming from an arbitrary direction}
\end{figure}

The metric tensor transforms like \cite{key-5,key-8}:

\begin{equation}
g^{ik}=\frac{\partial x^{i}}{\partial x'^{l}}\frac{\partial x^{k}}{\partial x'^{m}}g'^{lm}.\label{eq: trasformazione metrica}\end{equation}

Using eq. (\ref{eq: rotazione}), eq. (\ref{eq: rotazione 2}) and
eq. (\ref{eq: trasformazione metrica}), in the new rotated frame,
the line element (\ref{eq: metrica + 3}) in the $\overrightarrow{u}$
direction becomes (here we can neglect the $v$ and $w$ directions
because we will use bouncing photons and the photon deflection into
the $v$ and $w$ directions will be at most of order $h,$ then,
to first order in $h$, we can neglect the $dv^{2}$ and $dw^{2}$
terms):

\begin{equation}
ds^{2}=[1+(\cos^{2}\theta\cos^{2}\phi-\sin^{2}\phi)h^{+}(t-u\sin\theta\cos\phi)](du^{2}-dt^{2}).\label{eq: metrica + lungo u}\end{equation}

We emphasize that, in the line element (\ref{eq: metrica + lungo u})
a spatial dependence and an angular dependence are present in the
$u$ direction, differently from the line element of eq. 27 of \cite{key-7})
where, because of the simplest geometry there is a purely time dependence. 

The condition for null geodesics ($ds^{2}=0$) in eq. (\ref{eq: metrica + lungo u})
gives the coordinate velocity of the photon:

\begin{equation}
du^{2}=dt^{2}.\label{eq:u uguale t}\end{equation}

Thus, also in this case, the analysis of \cite{key-7} cannot start
directly from the condition of null geodesic. In fact also the metric
(\ref{eq: metrica + lungo u}) is not a {}``synchrony coordinate
system''. Thus also in the coordinates (\ref{eq: metrica + lungo u})
$t$ is only a time coordinate (i.e. the rate $d\tau$ of the proper
time is related to the rate $dt$ of the time coordinate from eq.
(\ref{eq: relazione temporale})). Then in this case a generalization
of our analysis of Section 4 can be applied like in Section 7 of \cite{key-5}.

From eq. (\ref{eq: metrica + lungo u}) it is

\begin{equation}
g_{00}=[1+(\cos^{2}\theta\cos^{2}\phi-\sin^{2}\phi)h^{+}(t-u\sin\theta\cos\phi)].\label{eq: gzz}\end{equation}
. 

Then, by using eq. (\ref{eq:u uguale t}) we obtain

\begin{equation}
d\tau^{2}=[1+(\cos^{2}\theta\cos^{2}\phi-\sin^{2}\phi)h^{+}(t-u\sin\theta\cos\phi)])du^{2},\label{eq: relazione spazial-temporale}\end{equation}

which gives 

\begin{equation}
d\tau=\pm[1+(\cos^{2}\theta\cos^{2}\phi-\sin^{2}\phi)h^{+}(t-u\sin\theta\cos\phi)]{}^{\frac{1}{2}}du.\label{eq: relazione temporale 2}\end{equation}

Now it will be shown that the analysis of \cite{key-7} works in this
case too.

We put the beam splitter in the origin of the new coordinate system
(i.e. $u_{b}=0$, $v_{b}=0$, $w_{b}=0$) \cite{key-5}. From eqs.
(\ref{eq: moto sol Corda}) it is known that an inertial test mass
initially at rest in the $x-y$ plane in our coordinates, remains
at rest throughout the entire passage of the GW. Eqs. (\ref{eq: moto sol Corda})
also show that the coordinates of the beam-splitter and of the mirror
change under the influence of the GW in the $z$ direction, but this
fact does not influence the total variation of the round trip proper
time of the photon (eq. (\ref{eq: tempo proprio totale lungo z in Corda})).
Then, in the computation of the variation of the proper distance in
our coordinates, the coordinates of the beam-splitter $u_{b}=0$ and
of the mirror $u_{m}=L$ can be considered fixed even in the $u-v$
plane, because the rotation (\ref{eq: rotazione}) does not change
the situation. Thus the proper duration of the forward trip can be
found as

\begin{equation}
\tau_{1}(t)=\int_{0}^{L}[1+(\cos^{2}\theta\cos^{2}\phi-\sin^{2}\phi)h^{+}(t-u\sin\theta\cos\phi)]{}^{\frac{1}{2}}du.\label{eq: relazione temporale 2}\end{equation}

with 

\begin{center}$t'=t-(L-u)$.\end{center}

In the last equation $t'$ is the retardation time (see Section 4
and refs. \cite{key-5,key-7}).

To first order in $h^{+}$ this integral can be approximated with

\begin{equation}
\tau_{1}(t)=T+\frac{\cos^{2}\theta\cos^{2}\phi-\sin^{2}\phi}{2}\int_{0}^{L}h^{+}(t'-u\sin\theta\cos\phi)du,\label{eq: durata volo andata approssimata u}\end{equation}

where

\begin{center}$T=L$ \end{center}

is the transit time of the photon in the absence of the GW. Similiary,
the duration of the return trip will be\begin{equation}
\tau_{2}(t)=T+\frac{\cos^{2}\theta\cos^{2}\phi-\sin^{2}\phi}{2}\int_{L}^{0}h^{+}(t'-u\sin\theta\cos\phi)(-du),\label{eq: durata volo ritorno approssimata u}\end{equation}

though now the retardation time is 

\begin{center}$t'=t-(u-l)$.\end{center}

The round-trip time will be the sum of $\tau_{2}(t)$ and $\tau_{1}[t-T_{2}(t)]$,
where $T_{2}(t)$ is the coordinate time which corresponds to $\tau_{2}(t)$.
The latter can be approximated by $\tau_{1}(t-T)$ because the difference
between the exact and the approximate values is second order in $h^{+}$.
Thus, to first order in $h^{+}$, the duration of the round-trip will
be

\begin{equation}
T_{r.t.}(t)=T_{1}(t-T)+T_{2}(t).\label{eq: durata round trip}\end{equation}

Using eqs. (\ref{eq: durata volo andata approssimata u}) and (\ref{eq: durata volo ritorno approssimata u})
it appears immediatly that deviations of this round-trip time (i.e.
proper distance) from its imperurbated value are given by

\begin{equation}
\begin{array}{c}
\delta T(t)=\frac{\cos^{2}\theta\cos^{2}\phi-\sin^{2}\phi}{2}\int_{0}^{L}[h^{+}(t-2T+u(1-\sin\theta\cos\phi))+\\
\\+h^{+}(t-u(1+\sin\theta\cos\phi))]du.\end{array}\label{eq: variazione temporale in u}\end{equation}

Now, with the auxilium of the Fourier transform of the $+$ polarization
of the field, defined by

\begin{equation}
\tilde{h}^{+}(\omega)=\int_{-\infty}^{\infty}dth^{+}(t)\exp(i\omega t)\label{eq: trasformata di fourier}\end{equation}
we obtain, in the frequency domain:

\begin{equation}
\delta\tilde{T}(t)=(\cos^{2}\theta\cos^{2}\phi-\sin^{2}\phi)\tilde{H}_{u}^{+}(\omega,\theta,\phi)\tilde{h}^{+}(\omega)\label{eq: segnale in frequenza lungo u}\end{equation}

where

\begin{equation}
\begin{array}{c}
\tilde{H}_{u}^{+}(\omega,\theta,\phi)=\frac{-1+\exp(2i\omega L)}{2i\omega(1-\sin^{2}\theta\cos^{2}\phi)}+\\
\\+\frac{\sin\theta\cos\phi((1+\exp(2i\omega L)-2\exp i\omega L(1+\sin\theta\cos\phi)))}{2i\omega(1-\sin^{2}\theta\cos^{2}\phi)},\end{array}\label{eq: fefinizione Hu}\end{equation}

and we immediately see that $\tilde{H}_{u}^{+}(\omega,\theta,\phi)\rightarrow L$
when $\omega\rightarrow0$.

Thus, the total response function of the arm of the interferometer
in the $\overrightarrow{u}$ direction to the $+$ component of the
GW is:

\begin{equation}
\Upsilon_{u}^{+}(\omega)=\frac{(\cos^{2}\theta\cos^{2}\phi-\sin^{2}\phi)}{2L}\tilde{H}_{u}^{+}(\omega,\theta,\phi)\tilde{h}^{+}(\omega)\label{eq: risposta + lungo u}\end{equation}

where $2L=2T$ is the round trip time in absence of gravitational
waves (note that in \cite{key-7} the Laplace transforms have been
used. Here we use the Fourier ones because the frequency response
functions of the Virgo interferometer for the two polarizations of
the GW will be designed, like in \cite{key-5} for the scalar case).

In the same way the line element (\ref{eq: metrica + 3}) in the $\overrightarrow{v}$
direction becomes:

\begin{equation}
ds^{2}=[1+(\cos^{2}\theta\sin^{2}\phi-\cos^{2}\phi)h^{+}(t-v\sin\theta\sin\phi)](dv^{2}-dt^{2}),\label{eq: metrica + lungo v}\end{equation}

and the response function of the $v$ arm of the interferometer to
the $+$ polarization of the GW will be:

\begin{equation}
\Upsilon_{v}^{+}(\omega)=\frac{(\cos^{2}\theta\sin^{2}\phi-\cos^{2}\phi)}{2L}\tilde{H}_{v}^{+}(\omega,\theta,\phi)\tilde{h}^{+}(\omega)\label{eq: risposta + lungo v}\end{equation}

where now it is

\begin{equation}
\begin{array}{c}
\tilde{H}_{v}^{+}(\omega,\theta,\phi)=\frac{-1+\exp(2i\omega L)}{2i\omega(1-\sin^{2}\theta\sin^{2}\phi)}+\\
\\+\frac{\sin\theta\sin\phi((1+\exp(2i\omega L)-2\exp i\omega L(1+\sin\theta\sin\phi)))}{2i\omega(1-\sin^{2}\theta\sin^{2}\phi)},\end{array}\label{eq: fefinizione Hv}\end{equation}

with $\tilde{H}_{v}^{+}(\omega,\theta,\phi)\rightarrow L$ when $\omega\rightarrow0$.

Thus the total response function (i.e. the angular frequency dependent
detector pattern) of an interferometer to the $+$ polarization of
the GW is:

\begin{equation}
\begin{array}{c}
\tilde{H}^{+}(\omega)=\frac{(\cos^{2}\theta\cos^{2}\phi-\sin^{2}\phi)}{2L}\tilde{H}_{u}(\omega,\theta,\phi)+\\
\\-\frac{(\cos^{2}\theta\sin^{2}\phi-\cos^{2}\phi)}{2L}\tilde{H}_{v}(\omega,\theta,\phi)\end{array}\label{eq: risposta totale Virgo +}\end{equation}

that in the low frequencies limit ($\omega\rightarrow0$) is in perfect
agreement with the detector pattern of eq. (\ref{eq: pattern piu'}):

\begin{equation}
\tilde{H}^{+}(\omega\rightarrow0)=\frac{1}{2}(1+\cos^{2}\theta)\cos2\phi.\label{eq: risposta totale approssimata}\end{equation}

Now the same analysis can be applied to the $\times$ polarization.
In this case the line element (\ref{eq: metrica polarizzazione per in gauge Corda})
has to be considered, and, using eq. (\ref{eq: rotazione}), eq. (\ref{eq: rotazione 2})
and eq. (\ref{eq: trasformazione metrica}), in the new rotated frame
the line element (\ref{eq: metrica polarizzazione per in gauge Corda})
in the $\overrightarrow{u}$ direction becomes:

\begin{equation}
ds^{2}=[1-2\cos\theta\cos\phi\sin\phi h^{\times}(t-u\sin\theta\cos\phi)](du^{2}-dt^{2}).\label{eq: metrica per  lungo u}\end{equation}

In this way the response function of the $u$ arm of the interferometer
to the $\times$ polarization of the GW is:

\begin{equation}
\Upsilon_{u}^{\times}(\omega)=\frac{-\cos\theta\cos\phi\sin\phi}{L}\tilde{H}_{u}^{\times}(\omega,\theta,\phi).\label{eq: risposta per lungo u}\end{equation}

In the analogous way the line element (\ref{eq: metrica polarizzazione per in gauge Corda})
in the $\overrightarrow{v}$ direction becomes:

\begin{equation}
ds^{2}=[1+2\cos\theta\cos\phi\sin\phi h^{\times}(t-u\sin\theta\sin\phi)](dv^{2}-dt^{2})\label{eq: metrica per  lungo v}\end{equation}

and the response function of the $v$ arm of the interferometer to
the $\times$ polarization of the GW is:

\begin{equation}
\Upsilon_{v}^{\times}(\omega)=\frac{\cos\theta\cos\phi\sin\phi}{L}\tilde{H}_{v}^{\times}(\omega,\theta,\phi)\label{eq: risposta per lungo v}\end{equation}

Thus the detector pattern of an interferometer to the $\times$ polarization
of the GW is:

\begin{equation}
\tilde{H}^{\times}(\omega)=\frac{-\cos\theta\cos\phi\sin\phi}{L}[\tilde{H}_{u}^{\times}(\omega,\theta,\phi)+\tilde{H}_{v}^{\times}(\omega,\theta,\phi)]\label{eq: risposta totale Virgo per}\end{equation}

that in the low frequencies limit ($\omega\rightarrow0$) is in perfect
agreement with the detector pattern of eq. (\ref{eq: pattern per}):

\begin{equation}
\tilde{H}^{\times}(\omega\rightarrow0)=-\cos\theta\sin2\vartheta.\label{eq: risposta totale approssimata}\end{equation}

Then it has been shown that, with the auxilium of our solution of
LEFEs in vacuum, a generalization of the analysis in \cite{key-7}
works in the computation of the two frequency - dependent detector
patterns of interferometers exactly like for the scalar waves in \cite{key-5}.

In figs. 4 and 5 the absolute value of the total response function
of the Virgo interferometer ($L=3$Km) for the $+$ and $\times$
polarizations of gravitational waves propagating from the direction
$\theta=\frac{\pi}{4}$ and $\phi=\frac{\pi}{3}$ are respectively
shown. From the figures it appears that at high frequencies the absolute
value of the response function decreases respect to the constant value
of the low frequencies approximation.

\begin{figure}
\includegraphics{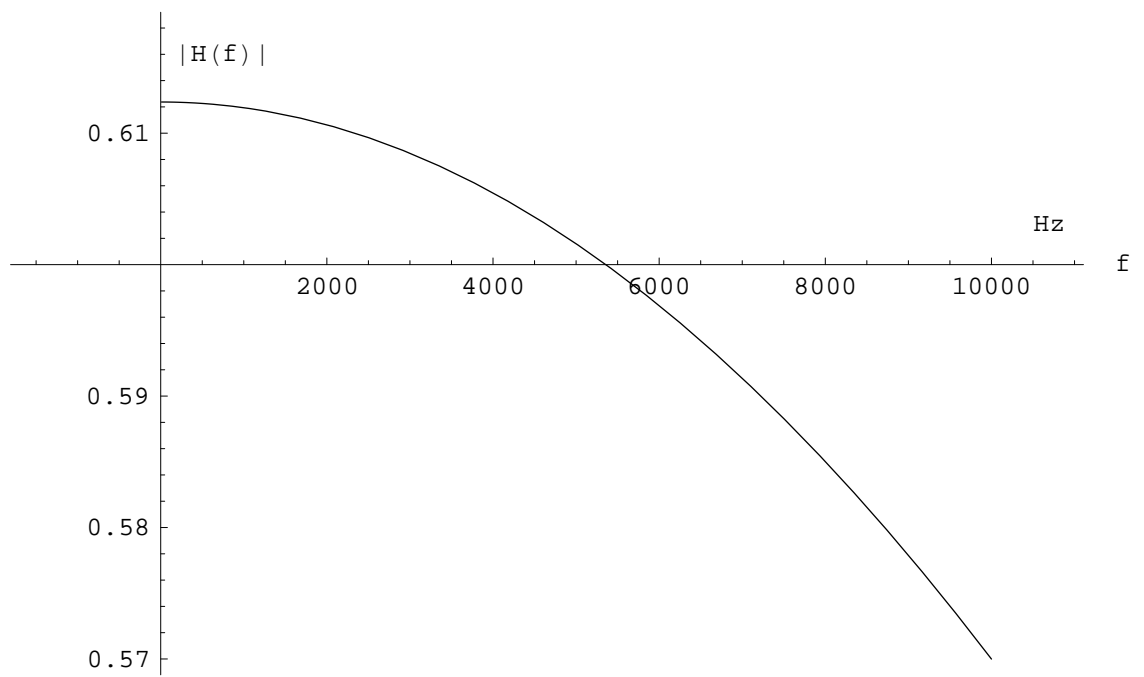}

\caption{the absolute value of the total response function of the Virgo interferometer
to the $+$ polarization of the gravitational waves for $\theta=\frac{\pi}{4}$
and $\phi=\frac{\pi}{3}$. }
\end{figure}
\begin{figure}
\includegraphics{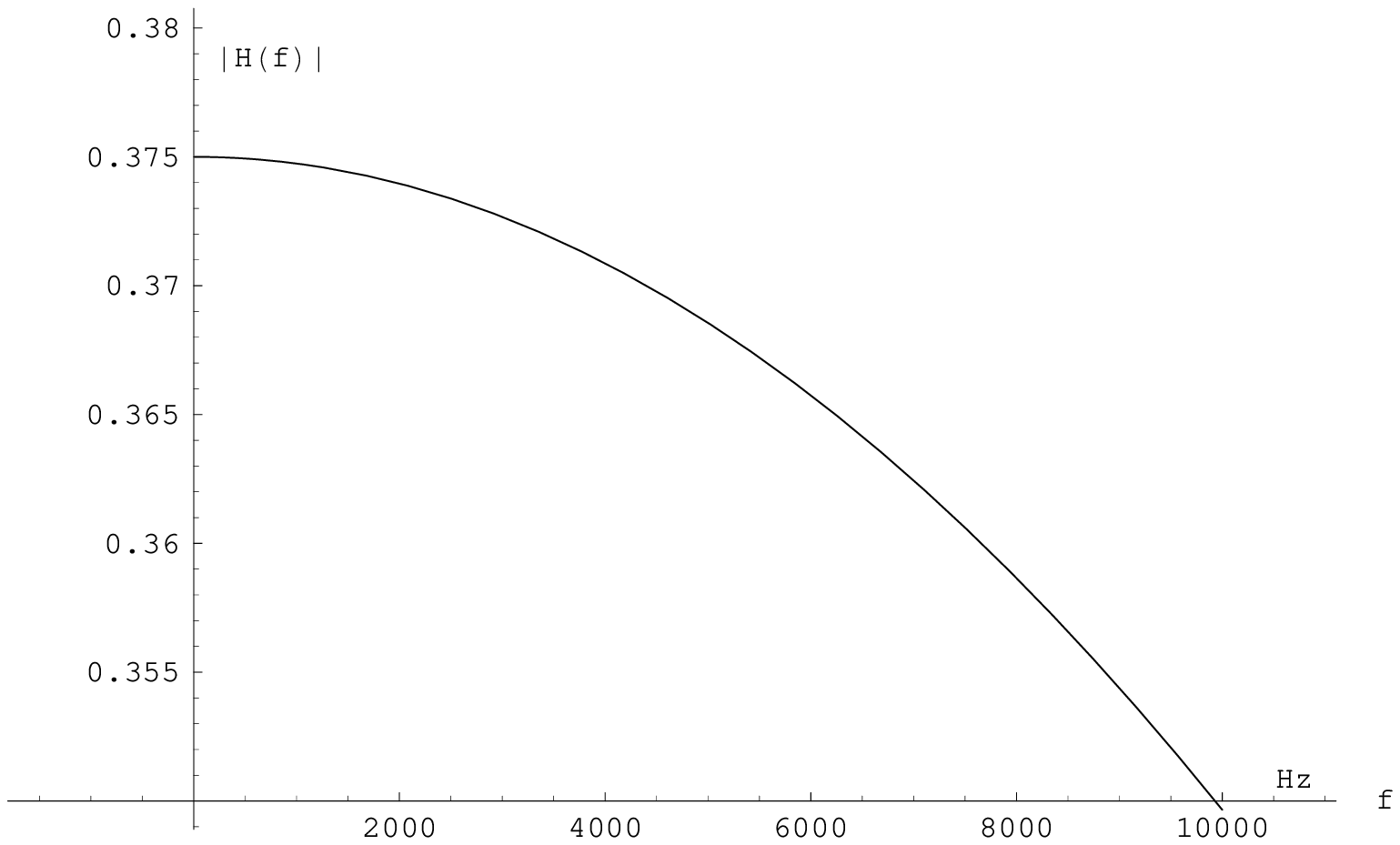}

\caption{the absolute value of the total response function of the Virgo interferometer
to the $\times$ polarization of the gravitational waves for $\theta=\frac{\pi}{4}$
and $\phi=\frac{\pi}{3}$. }
\end{figure}

\section{The total output due to the stochastic background of gravitational
waves}

With the auxilium of the frequency dependent pattern functions the
approximate eqs. (\ref{eq: formula  approssimata in frequenza2})
and (\ref{eq: legame output-stochastic}) can be generalized with,
respectively,

\begin{equation}
\tilde{S}(\omega)=\int_{S^{2}}d\hat{\Omega}[\tilde{H}^{\times}(\omega)\tilde{h}^{\times}+\tilde{H}^{+}(\omega)\tilde{h}^{+}]\label{eq: formula  esatta}\end{equation}

and

\begin{equation}
s(t)=\frac{1}{2\pi}\int_{-\infty}^{+\infty}d\omega\int_{S^{2}}d\hat{\Omega}[\tilde{H}^{\times}(\omega)\tilde{h}^{\times}+\tilde{H}^{+}(\omega)\tilde{h}^{+}]\exp(i\omega t).\label{eq: legame output-stochastic2}\end{equation}

\section{Conclusions }

A solution of linearized Einstein field equations in vacuum has been
given and discussed. First  it has been shown that, if from the metric
(\ref{eq: metrica + 3}) the linearized connections, the linearized
Riemann tensor and the linearized Ricci tensor are computed, the linearized
Ricci tensor is equal to zero (the metric (\ref{eq: metrica + 3})
is a solution of LEFEs in vacuum). Then the effect on test masses
of the particular solution (\ref{eq: metrica + 3}), which is a gravitational
wave has been discussed. It has been shown that in our solution test
masses have an apparent motion in the direction of propagation of
the wave, while in the transverse direction they appear at rest. In
this way it could appear that GWs would be longitudinal waves, but,
from careful investigation of the solution (\ref{eq: metrica + 3}),
it has been found that the tidal forces associated with GWs act along
the directions orthogonal to the direction of propagation of waves.
The computation has first been made in the long wavelenghts approximation
(wavelength much larger than the linear dimensions of the interferometer),
then the analysis has been applied to all GWs using a generalization
to our solution of the analysis of \cite{key-5,key-7}.

After this, in the context of the potential detection of stochastic
backgrounds of gravitational waves with interferometers, which is
very important in a cosmological scenario because a stochastic background
of GWs could be the result of processes that took place in a time
very close to the Planck era, our solution for LEFEs in vacuum has
been used to generalize to all the frequencies the concept of detector
pattern of interferometers, which was well known in the assumption
that the wavelength of the GWs is much larger than the distance between
the test masses (i.e. the beam-splitter and the mirrors of the interferometer).
In the low frequencies approximation our results agree with the standard
detector patterns computed in in lots of works in literature (see
for example \cite{key-9,key-12,key-13,key-14,key-15}), while, at
high frequencies, it has been obteined that the absolute value of
the total response function of interferometers decreases with respect
to the constant value of the long wavelenghts approximation.

We emphasize that, also in the analysis of the angular pattern of
interferometers, the analysis of \cite{key-5,key-7} has been generalized:
in \cite{key-7} only the simplest geometry case in the TT solution
was considered, while in \cite{key-5} the analysis worked for purely
scalar waves. It has been shown that the analysis can be generalized
to the more general geometry and to our solution for LEFEs in vacuum
too. 

At the end of this paper we have computed the total signal seen from
an interferometer for the stochastic background of gravitational waves.

\section*{Acknowledgements}

I would like to thank Mariafelicia De Laurentis and Mauro Francaviglia
for helpful advices during my work. The European Gravitational Observatory
(EGO) consortium has also to be thanked for the using of computing
facilities.

\end{document}